\documentclass[pra,reprint,a4paper,twocolumn,nofootinbib,showpacs,superscriptaddress,showkeys]{revtex4-1}
\usepackage{amsmath,amssymb,amsfonts}
\usepackage[dvips]{graphicx}
\usepackage{graphicx}
\usepackage{txfonts}
\usepackage[T1]{fontenc}
\usepackage{textcomp}
\usepackage{graphicx}
\usepackage{epsfig}
\usepackage{upgreek}
\usepackage{braket}
\usepackage{bbold}
\usepackage{psfrag}
\usepackage[mathscr]{euscript}
\usepackage[unicode]{hyperref}
\hypersetup{
    unicode=true,                                           
    a4paper=true,
    plainpages=false,
    pdffitwindow=true,                                      
    pdftitle={Soliton Interferometry},                      
    pdfauthor={John Lloyd Helm},                            
    pdfsubject={},                                          
    pdfkeywords={Bose--Einstein, condensate, BEC, solitons},
    colorlinks=true,                                        
    linkcolor=blue,                                         
    citecolor=blue,                                         
    filecolor=blue,                                         
    urlcolor=blue                                           
}
\newcommand{\eqnrefp}[1]{{[Eq.~(\ref{#1})]}}

\newcommand{\eqnreft}[1]{{Eq.~(\ref{#1})}}

\newcommand{\figreft}[2]{Fig.~\ref{#1}#2}

\newcommand{\figssssreft}[4]{Figs.~\ref{#1},~\ref{#2},~\ref{#3} and \ref{#4}}
\newcommand{\figssreft}[3]{Figs.~\ref{#1}#2 and #3}
\newcommand{\figreftfull}[2]{Figure~\ref{#1}#2}
\newcommand{\figrefp}[2]{[Fig.~\ref{#1}#2]}

\newcommand{\secreft}[1]{{Section~\ref{#1}}}
\newcommand{\secrefp}[1]{{(Sec.~\ref{#1})}}

\newcommand{\ans}{\emph{Ansatz}}
\renewcommand{\H}{\mathscr{H}}
\newcommand{\HID}{\mathscr{H}_{\scriptsize{\mbox{1D}}}}
\newcommand{\HOD}{\mathrm{H}_{\scriptsize{\mbox{1D}}}}
\newcommand{\POD}{\Psi_{\scriptsize{\mbox{1D}}}}
\newcommand{\ES}{E_{\scriptsize\mbox{S}}}
\newcommand{\EK}{E_{\scriptsize\mbox{K}}}
\newcommand{\EG}{E_{\scriptsize\mbox{G}}}
\newcommand{\EB}{E_{\scriptsize\mbox{B}}}
\newcommand{\eye}{i}
\newcommand{\pa}{\partial}
\newcommand{\xp}{\,\mathrm{e}^}

\newcommand{\xc}{x_{\scriptsize\mathrm{C}}}
\newcommand{\kc}{k_{\scriptsize\mathrm{C}}}

\newcommand{\HC}{H_{\scriptsize\mathrm{C}}}
\newcommand{\Sx}{s_{x}}
\newcommand{\Sv}{s_{v}}
\newcommand{\Sk}{s_{k}}
\newcommand{\xf}{x_{\mathrm{f}}}
\newcommand{\vf}{v_{\mathrm{f}}}
\newcommand{\vb}{v_{\mathrm{b}}}
\newcommand{\mvb}{\mu_{v_{\mathrm{b}}}}
\newcommand{\svb}{\sigma_{v_{\mathrm{b}}}}
\newcommand{\HR}{H_{\scriptsize\mathrm{R}}}
\newcommand{\PC}{\psi_{\scriptsize\mathrm{C}}}
\newcommand{\FC}{\phi_{\scriptsize\mathrm{C}}}

\newcommand{\E}[1]{\mathrm{E}\left[#1\right]}
\newcommand{\V}[1]{\mathrm{Var}\left[#1\right]}
\newcommand{\inp}[1]{\langle#1\rangle}

\newcommand{\Tom}{\omega_{\scriptsize\mathrm{T}}}
\newcommand{\fomega}{\omega_{\scriptsize\mathrm{f}}}
\newcommand{\xom}{\omega_x}
\newcommand{\Mdelta}{\delta_{\scriptsize\mathrm{MZ}}}

\newcommand{\sech}{\mathrm{sech}}
\renewcommand{\sin}{\mathrm{sin}}
\renewcommand{\cos}{\mathrm{cos}}

\begin{document}

\title{Splitting bright matter-wave solitons on narrow potential barriers: quantum to classical transition and applications to interferometry.}
\author{J. L. Helm}
\affiliation{Joint Quantum Centre Durham-Newcastle, Department of Physics, Durham University, Durham DH1 3LE, United Kingdom}
\author{S. J. Rooney}
\affiliation{Jack Dodd Centre for Quantum Technology, Department of Physics, University of Otago, Dunedin, New Zealand}
\author{Christoph Weiss}
\affiliation{Joint Quantum Centre Durham-Newcastle, Department of Physics, Durham University, Durham DH1 3LE, United Kingdom}
\author{S. A. Gardiner}
\affiliation{Joint Quantum Centre Durham-Newcastle, Department of Physics, Durham University, Durham DH1 3LE, United Kingdom}
\date{\today}

\begin{abstract}
We study bright solitons in the Gross--Pitaevskii equation as they are split and recombined in a low energy system. We present new analytic results determining the general region in which a soliton may not be split on a potential barrier, and confirm these results numerically. Furthermore, we analyse the energetic regimes where quantum fluctuations in the initial center of mass position and momentum become influential on the outcome of soliton splitting and recombination events. We then use the results of this analysis to determine a parameter regime where soliton interferometry is practicable. 
\end{abstract}

\maketitle
\section{Introduction\label{intro}}

Atomic Bose-Einstein condensates (BECs) with attractive inter-atomic interactions are capable of supporting soliton-like dynamical excitations referred to as bright solitary matter-waves\cite{khaykovich_etal_science_2002, strecker_etal_nature_2002, cornish_etal_prl_2006, Marchant_etal_2013, Hulet2010b}.  These excitations are soliton-like in the sense that they propagate without dispersion \cite{morgan_etal_pra_1997}, are robust to collisions with both other bright solitary matter-waves and slowly varying external potentials \cite{parker_etal_physicad_2008, billam_etal_pra_2011}, and have center-of-mass trajectories which are well-described by effective particle models \cite{martin_etal_prl_2007, martin_etal_pra_2008, poletti_etal_prl_2008}. They derive these soliton-like properties from their analogousness to the bright soliton solutions of the focusing nonlinear Schr\"{o}dinger equation (NLSE)~\cite{zakharov_shabat_1972_russian, satsuma_yajima_1974, gordon_ol_1983, haus_wong_rmp_1996, Helczynski_ps_2000}, to which the mean-field description of an atomic BEC reduces in an effectively unconfined, quasi-one-dimensional (quasi-1D) limit. Although the quasi-1D limit is experimentally challenging for attractive condensates \cite{billam_etal_variational_2011}, bright solitary matter-wave dynamics remain highly soliton-like outside this limit \cite{cornish_etal_prl_2006, billam_etal_pra_2011}. Consequently, bright solitary matter-waves present an intriguing candidate system for future interferometric devices \cite{HELM_PRA_2012,strecker_etal_nature_2002, cornish_etal_physicad_2009, weiss_castin_prl_2009, streltsov_etal_pra_2009, billam_etal_pra_2011, al_khawaja_stoof_njp_2011, martin_ruostekoski_njp_2010,busch_pra_2013,cuevas_njp_2013}.

The collision of a bright solitary wave with a narrow potential barrier is a good candidate for a mechanism for the creation of coherent localised condensates, much as a beamsplitter coherently splits a light beam in an optical interferometer. This mechanism has been investigated extensively in the quasi-1D, mean-field description of an atomic BEC~\cite{HELM_PRA_2012,kivshar_malomed_rmp_1989, ernst_brand_pra_2010,lee_brand_2006, cao_malomed_pla_1995, holmer_etal_cmp_2007, holmer_etal_jns_2007,POLO_etal_PRA_2013,WangEtAl2012,Molmer_arxiv_2012,Minmar_thesis_2012}, and sufficiently fast collisions with potential barriers have been shown to lead to the desired beamsplitting effect \cite{holmer_etal_cmp_2007, holmer_etal_jns_2007}. Similarly, the dynamics of solitons has been studied in nonlinear optics in an inhomogeneous array of discrete waveguides. In this system the inhomogeneity facilitates reflection, splitting or capture  of the soliton~\cite{Krolikowski_joptsocam_1996, Fratalocchi_pre_2006, Konotop_PRE_1996}. This is equivalent, in the continuum limit of an infinite number of waveguides, to splitting a soliton in the Gross-Pitaevskii equation (GPE) at a $\delta$-function potential barrier~\cite{Krolikowski_joptsocam_1996}. In the optics community this phenomenon has been called the ``optical axe''~\cite{Helczynski_ps_2000}. Incomplete/bound state splitting has been considered in the context of soliton molecule formation \cite{al_khawaja_stoof_njp_2011}, within a mean-field description, and also in the context of many-body quantum mechanical descriptions: in the latter it has been demonstrated that macroscopic quantum superpositions of solitary waves could be created, offering intriguing possibilities for future atom interferometry experiments \cite{weiss_castin_prl_2009, streltsov_etal_pra_2009}.  

A related work~\cite{martin_ruostekoski_njp_2010} considered an interferometer using a narrow potential barrier as a beamsplitter for harmonically trapped solitary waves, based on the particular configuration of a recent experiment ~\cite{nlqugas_confproc_2010}. In particular this work demonstrated that such a potential barrier can also be used to recombine solitary waves, by arranging for them to collide at the location of the barrier. The dynamics of these collisions were further explained in Ref.~\cite{HELM_PRA_2012}. In such collisions, the relative norms of the two outgoing solitary waves was shown to be governed by the phase difference $\Delta$ between the incoming ones.  In the mean-field description the relative norms of the outgoing waves exhibit enhanced sensitivity to small variations in the phase $\Delta$; however, a simulation of the same system including quantum noise via the truncated Wigner method \cite{blakie_etal_ap_2008}, showed increased number fluctuations that ultimately negated this enhancement \cite{martin_ruostekoski_njp_2010}.

In the current work, our first result will be to carefully explore the spectrum of splitting behaviours which these systems can exhibit. It has been established that quantum superpositions, in the form of ``NOON states'' or ``Schr\"{o}dinger cat states'' can be created when the energy associated with the splitting event is particularly low~\cite{weiss_castin_prl_2009, Gertjerenken_etal_pra_2012}. Here, we wish to determine the location of the boundary between this quantum behaviour and more classical behaviour, which will determine where interferometry is a more practical goal. We will also present a rigorous determination of the phase shift accrued between the resulting solitons after a splitting event, based on the work presented in~\cite{holmer_etal_cmp_2007}. Our second major result will be to more thoroughly outline two different geometries which might be employed for soliton interferometry, and again delineate energetic regimes where these implementations are practicable.

The current publication is presented as follows: In~\secreft{sec:system} we formally introduce the 3D mean-field Hamiltonian of the system, the reduced 1D Hamiltonian and the associated dynamic equation (the GPE). In~\secreft{sec:soliton_splitting} we outline the energetic regimes of soliton splitting in the GPE, presenting analytic results in~\secreft{sec:class_analysis} and comparing these results to numerical simulations in~\secreft{sec:class_num}. We then establish the quantum uncertainties associated with the harmonically trapped system~\secrefp{sec:quant_analysis}. These uncertainties are used to determine a sensitivity measure of the equal splitting case~\secrefp{sec:trans_sens} and the distributions of the transmission after the addition of quantum fluctuations~\secrefp{sec:montecarlo_split}. The last part of this section,~\secreft{sec:split_shift}, presents a derivation of the split induced phase shift. The final results section~\secrefp{sec:interferometry} outlines how these results might be implemented to perform Mach-Zehnder interferometry in a torus~\secrefp{sec:tor_mac} and Mach-Zehnder interferometry in a harmonic trap~\secrefp{sec:harm_mac}. In these sections we delimit regimes where these forms of interferometry are experimentally viable in terms of the collisional energy of the system. We also outline the effects of quantum uncertainty on the harmonically trapped interferometry case (\secreft{sec:sens_int}~and~\secreft{sec:montecarlo_int}). 

\section{Physical system\label{sec:system}}
We begin with the 3D $N$-particle mean-field energy Hamiltonian $\H[\psi]$ for a Bose field, defined as~\cite{pethick_smith_2002}
  \begin{align}
      \H[\Psi]=\int\mathrm{d}{\bf r}\Bigg[\frac{\hbar^2}{2m}|\nabla\Psi({\bf r})|^2 + &V_\mathrm{ext}({\bf r})|\Psi(\bf r)|^2\notag\\
                        &-\frac{2\pi N|a_s|\hbar^2}{m}|\Psi({\bf r})|^4\Bigg].
  \end{align}
Here $N$, $m$ and $a_s$ are the atom number, mass, and $s$-wave scattering length respectively. A delta function contact potential is assumed. For attractive inter-atomic interactions $a_s<0$. The wave function, $\Psi$, is normalised to 1. The potential $V_\mathrm{ext}(\mathbf{r})$ is comprised of both the trapping potentials and any external potentials used to construct narrow barriers used for splitting the soliton. We model this potential as
  \begin{equation}
    V_\mathrm{ext}({\bf r})=\EB\xp{-2x^2/x_r^2}+\frac{m}{2}\left[\Tom^2x^2+\omega_r\left(y^2+z^2\right)\right].\label{potential}
  \end{equation}
The first term describes the narrow splitting barrier and can be generated by an off-resonant Gaussian light sheet~\cite{Marchant_etal_2013} perpendicular to the $x$ direction with $1/e^2$ radius $x_r$ in the $x$ direction, with peak beam strength $\EB$. The second term denotes a standard magnetic harmonic confinement which we take to be a cylindrically symmetric waveguide; such a configuration is approximately achieved in an atomic waveguide trap.

By increasing the radial trapping we can reach a quasi-1D regime, as defined in detail in Ref.~\cite{billam_etal_variational_2011}, where the radial trapping is tight but remains 3D [$a_s\ll(\hbar/m\omega_r)^{1/2}$]. In this regime we can separate the radial and axial dynamics with the \ans~$\Psi(\mathbf{r})=\POD(x) (m\omega_r/\pi\hbar)^{1/2}\exp{(-m\omega_r[y^2+z^2]/2\hbar)}$. After factoring out global phases associated with the radial harmonic ground state energies this yields both the quasi-1D classical field Hamiltonian~\cite{pethick_smith_2002},
  \begin{align}
      \HID[\POD]=\int\mathrm{d}{x}\Bigg[\frac{\hbar^2}{2m}\left|\frac{\pa}{\pa x}\POD(x)\right|^2 + V_{\scriptsize{\mbox{ext}}}(x)&|\POD(x)|^2\notag\\
                        -\frac{gN}{2}&|\POD(x)|^4\Bigg],
  \label{eqn:HID}
  \end{align}
and its associated quasi-1D GPE~\cite{pethick_smith_2002}
   \begin{equation}
     \eye\hbar\frac{\pa \POD(x)}{\pa t}=\left[-\frac{\hbar^2}{2m}\frac{\pa^2}{\pa x^2}+ V_\mathrm{ext}(x) -gN\left|\POD(x)\right|^2\right]\POD(x).
   \end{equation}
The non-linearity is quantified by $g=2\hbar \omega_r |a_s|$. If we take $V_\mathrm{ext}=0$ then this equation reduces to the NLSE. We will also consider a toroidal ring trap~\cite{neely_prl_2013,ryu_prl_2013,Ramanathan_etal_prl_2011}   by setting $\Tom=0$ and introducing periodicity in x.

Working in soliton units ---  position units of $\hbar^2/mgN$, time units of $\hbar^3/mg^2N^2$, and energy units of $mg^2N^2/\hbar^2$ \cite{billam_etal_variational_2011} --- yields the dimensionless, quasi-1D GPE\footnote{It should be noted that in the very low $N$ limit this rescaling takes a slightly different form, with $N$ replaced by $N-1$. This rescaling is used in Ref.~\cite{Holdaway_etal_2012}}
  \begin{multline}
    \eye\frac{\pa \psi(x)}{\pa t} =\Bigg[- \frac{1}{2}\frac{\pa^2}{\pa x^2} + \frac{q}{\sigma_\mathrm{b}\sqrt{2\pi}}e^{-x^2/2\sigma^2}\\
    +\frac{\xom^2x^2}{2}-\left|\psi(x)\right|^2\Bigg]\psi(x),
    \label{eqn:1DGPE}
  \end{multline}
where the dimensionless wave function is $\psi = \hbar\POD/\sqrt{mgN}$, the barrier width is characterised by $\sigma_\mathrm{b}$ (the dimensionless form of half the $1/e^2$ radius) and the barrier strength is given by 
  \begin{equation}
    q=\sqrt{\frac{\pi}{2}}\frac{\EB x_r}{gN}.
  \end{equation}
  
\begin{figure*}[t]
    \includegraphics[width=\textwidth]{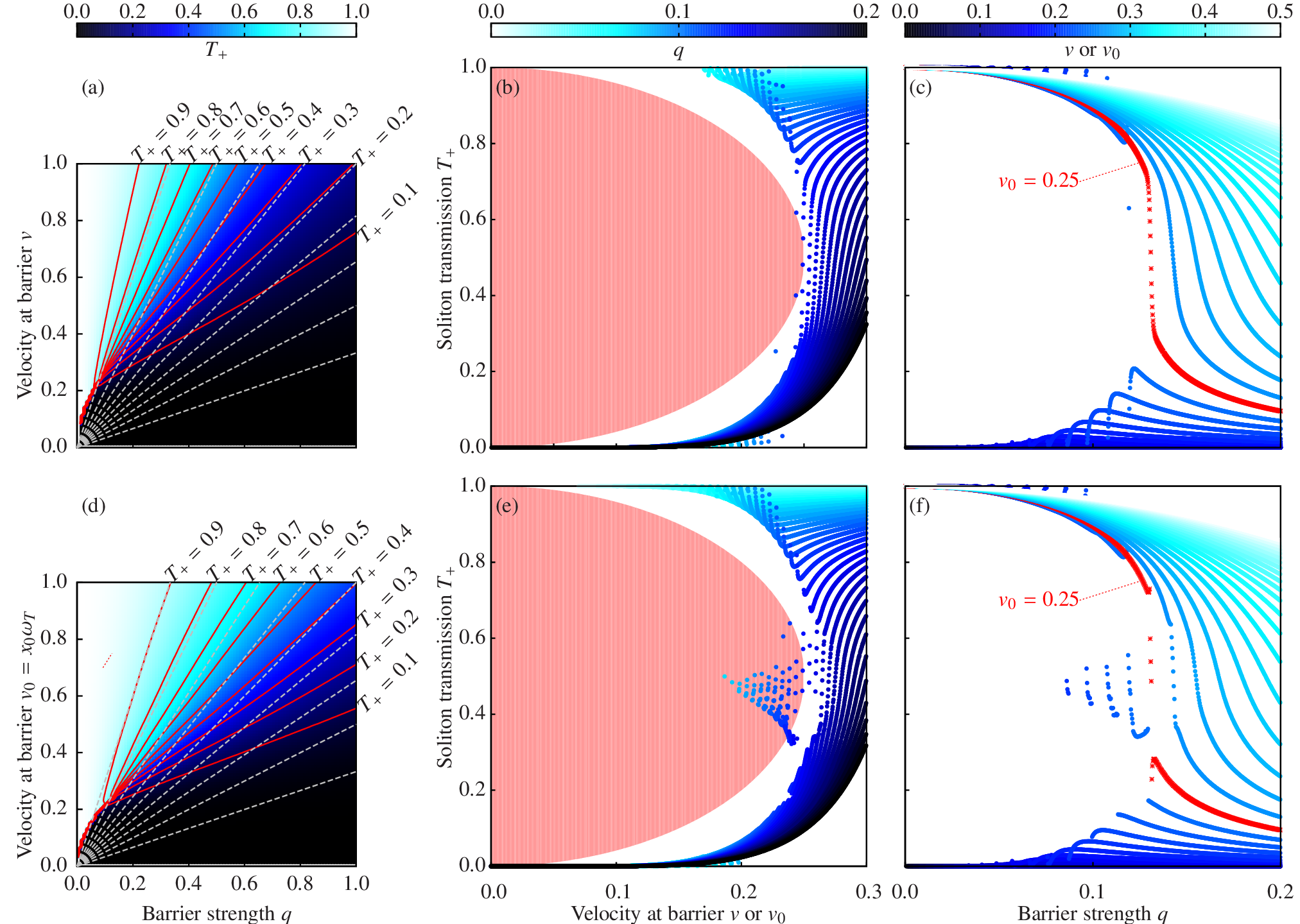}
    \caption{(Color online) Numerical results of splitting a soliton travelling at velocity $v$-(a-c) or $v_0$-(d-f) at a Gaussian barrier of strength $q$ and width $\sigma=0.2$. (a,d) Colormaps of transmission as a function of $q$ and $v$ or $v_0$. The solid (red) curves are iso-lines of constant transmission $T_+$ obtained from the numerics, while the dashed (gray) curves are theoretical predictions of transmission $T_q^s$ in the linear case over the same range. (b,e) Curves of transmission as a function of collisional velocity $v$ or $v_0$ for various barrier strength $q$. The shaded (red) region shows energetically disallowed splitting events. (c,f) Curves of transmission as a function of barrier strength $q$ for various values of $v$ or $v_0$. The labelled (red) curve, for which $v,v_0=0.25$ indicates the classical, untrapped lower energy bound on the region where a continuous range of transmission is accessible.}
    \label{fig:pscan}
  \end{figure*}

\section{Soliton splitting\label{sec:soliton_splitting}}
\subsection{Overview\label{sec:soliton_splitting_overview}}

In this section we probe the transition from low- to high-energy soliton splitting. We quantify the energy by the velocity of the soliton at the point of collision with the barrier, denoted as: $v$ for an untrapped system, where the velocity is brought about by an imprinted phase on the initial condition; or $v_0$ for the axially trapped system, where the velocity is a result of the axial trapping $\xom$ being greater than zero and an initial offset $x_0$ in the initial condition. This offset separates the soliton from the point where the soliton is split by the barrier at $x=0$.

We take $v,v_0\gtrsim1.0$ to be the high energy regime and $v,v_0\lesssim0.25$ to be the low energy regime~\cite{Gertjerenken_etal_pra_2012}. As such, the transitional energy regime lies within the $0.25\le v,v_0\le1$ velocity range. We will justify the lower bound of this regime by considering classical descriptions of the kinetic and ground state energies of the system. We will also show that these arguments describe a process which is analogous to the quantum mechanical transition from product state wave-functions (where, after scattering the transmitted/reflected portions of the wave function can range continuously between zero and full transmission/reflection) to bimodal systems (where the soliton is either reflected by or transmitted through the barrier, but never split)\footnote{It should be noted that even in the high energy regime we cannot make a soliton of arbitrary size by simply scattering a larger soliton off a barrier. The scattered portion of the wavefunction may be too small to form a soliton and must be considered radiation~\cite{holmer_etal_cmp_2007}.}.

\subsection{Analysis of classical soliton splitting~\label{sec:class_analysis}}

We explain the transition between high and low energy dynamics by comparing the incoming collisional kinetic energy $\EK$ and the energy required to split the soliton $\ES$. Firstly, rescaling the quasi-1D Hamiltonian~\eqnrefp{eqn:HID} in to soliton units with $\Tom=0$ gives 
  \begin{equation}
      \HOD[\psi]=\int\mathrm{d}{x}\Bigg[\frac{1}{2}\left|\frac{\pa}{\pa x}\psi(x)\right|^2-\frac{1}{2}|\psi(x)|^4\Bigg].
  \label{eqn:H}
  \end{equation}
We then substitute the 1D soliton solution, 
  \begin{equation}
    \psi_0=\frac{1}{2}\sech\left(\frac{x-x_0}{2}\right)\xp{\eye vx},\label{eqn:PODSECH}
  \end{equation}
into our Hamiltonian, with $v=0$, and obtain both the per-particle soliton ground state energy ($\HOD[\psi_0]=-1/24$) and $N$-particle soliton ground state energy [$\EG(N)=-N/24$]. We then consider an $n$ particle soliton which is spatially well separated from the rest of the condensate and any potentials. Failure to satisfy this separation assumption may result in a bound state, and further contributions to the ground state energy will arise. The effects of such bound states will be discussed later. Assuming that the whole condensate contains a total of $N$ particles we see that the spatially separated soliton's contribution to the total energy is 
  \begin{equation}
    \EG(n)=-\frac{n}{24}\left(\frac{n}{N}\right)^2.
  \end{equation}
We reach this conclusion by rescaling the $n$-particle soliton ground state energy $\EG(n)$ into $N$ particle soliton units. This is equivalent to multiplying by $(n/N)^2$. By constructing the energy difference $\ES$ we can easily see that the energy required to split the soliton is 
  \begin{align}
       \ES&=\EG(N-n)+\EG(n)-\EG(N)\notag,\\
       &=3|\EG(N)|\left(1-\frac{n}{N}\right)\frac{n}{N}\label{eqn:ESEG}.
  \end{align}
We can now re-cast this result in terms of the transmission, $T_+$:
  \begin{equation}
    T_+=\int_0^\infty|\psi|^2\mathrm{d} x
       =\frac{n}{N}
  \end{equation}
yielding
  \begin{align}
    \ES&=\frac{1}{8}T_+\left(1-T_+\right)N.
  \end{align}
Next, we describe the classical particle energy of an $N$-particle soliton moving at velocity $v$:
  \begin{equation}
    \EK=\frac{v^2N}{2}.\\
  \end{equation}
We can now see that, for splitting to occur, we must satisfy $\EK>\ES$ and so
  \begin{equation}
    |v|>\frac{1}{2}\sqrt{T_+\left(1-T_+\right)}.\label{eqn:VSCOND}
  \end{equation}
This inequality describes the high energy regime in that parameters which do not satisfy it are only available in the low energy regime. If we consider the functional form of our inequality we see that $\sqrt{T_+\left(1-T_+\right)}$ is maximal for $T_+=0.5$, at which value we have $|v|>0.25$. As such, the first state to become inaccessible is the equal splitting case, which cannot be accessed for $|v|<0.25$. Equivalently, we must satisfy $\EK/\ES>0.75$~\eqnrefp{eqn:ESEG}. This is consistent with results described in Ref.~\cite{Gertjerenken_etal_pra_2012}.

As noted above, splitting the soliton reduces the amount of kinetic energy available to the solitons. In the high energy regime, this reduction is negligible and the solitons are capable of becoming well separated from the barrier, and one another, after the split occurs. At lower energies this is not always the case. As less and less energy is available to the resulting solitons their outgoing velocities are notably reduced, and eventually the solitons become trapped at the barrier. The effect of the harmonic trap enhances this effect, as the outgoing velocity determines the maximal separation which the resulting solitons can achieve. This phenomenon is shown in~\figreft{fig:pscan}~and will be discussed in the next section.

\subsection{Numerical analysis of classical soliton splitting~\label{sec:class_num}}

We numerically verify these results by evolving the initial condition described by~\eqnreft{eqn:PODSECH} according to the dynamics of~\eqnreft{eqn:1DGPE}. We perform two types of evolution. For the first type we set $\xom=0$ and perform integrations over a range of $v$ and $q$. These calculations allow us to consider the behaviour of the untrapped, true, soliton to which the above analytic results apply exactly. \figreft{fig:pscan}{(a-c)} shows the results of these simulations. For the second type of simulation we set the initial velocity $v=0$ and integrate over a range of $\xom$ and $q$. By keeping the initial offset constant at $x_0=-L/4$, where the numerical algorithm has spatial domain $-L/2<x\le L/2$, we are able to use $\xom$ to select a collisional velocity $v_0=\xom x_0$. This allows us to more accurately describe the behaviour we would see in an experiment where the soliton is accelerated by an axial harmonic trap. \figreft{fig:pscan}{(d-f)} shows the results of these equations. 

For all simulations the barrier is situated at the trap minimum (specifically $x=0$) and we set the barrier width to $\sigma_\mathrm{b}=0.2$. Barrier potentials of finite width/height have some limitations in the extremely high velocity regime, in that if the peak energy of the barrier is not notably higher than the kinetic energy of the soliton then the soliton classically passes over the barrier and no splitting occurs~\citep{HELM_PRA_2012} . This restricts the width of the barrier in a given energy regime by requiring that the barrier be narrow enough to constitute a quickly varying potential when compared to the incoming velocity of the soliton. The energy regimes we consider in the current work are compatible with a barrier width of $\sigma_{\mathrm{b}}\lesssim0.2$. A broader discussion of the effect of finite width (for a Rosen--Morse i.e. $\sech^2$ potential barrier) is presented in Ref.~\cite{POLO_etal_PRA_2013}.

\figreftfull{fig:pscan}{(a)} displays a broad scan of the $q,v$ parameter space. At higher velocities ($v>0.25$) we see a continuous range of transmissions is accessible. At lower velocities this is not the case, and for $v\lesssim0.1$ we see that we are effectively left with only full transmission and full reflection as accessible final states. 

We have displayed two sets of curves of constant transmission on~\figreft{fig:pscan}{(a)}: solid (red) and dashed (grey). The solid (red) curves are iso-lines of constant transmission $T_+=0.1, 0.2, \ldots 1.0$ taken from the colormap itself. At higher values of $v$ these curves are well separated, illustrating that we can access the full range of transmissions by selecting $q$ and $v$ accordingly. As $v$ decreases these curves begin to converge. The convergence of iso-curves signifies that the splitting state associated with the curves has become disallowed.

We derived the second set of curves in ~\figreft{fig:pscan}, the dashed (grey) curves, from analysis presented in~\cite{holmer_etal_cmp_2007}. The analysis states that for a $\delta$-function barrier in the regime where both the mean-field interpretation is valid and the velocity is high the transmission is given by
  \begin{align}
    T_q^s(v)&=\lim_{t\rightarrow \infty} \int_0^\infty|\psi(x,t)|^2\notag dx\\
            &=|t_q(v)|^2 =\frac{v^2}{v^2+q^2}=\frac{1}{1+\alpha^2}\label{eqn:transanalysis}.
  \end{align}
This analysis illustrates that in the high energy regime the transmission is determined solely by the ratio $\alpha=q/v$, and we predict the dashed (grey) curves of constant transmission which take the form
\begin{equation}
v=\left(\frac{T_q^s}{1-T_q^s}\right)^{1/2}q\label{eqn:hight}.
\end{equation}
Here we have adopted Holmer's $T_q^s$ notation to denote the limiting case of a high energy mean-field soliton colliding with a $\delta$-function barrier. In~\figreft{fig:pscan}~we display the curves for $T_q^s=0.1, 0.2, \ldots 1.0$. It should be noted that these curves are also the transmission rates of plane-waves though a $\delta$-function barrier in the linear Schr\"odinger equation, where the energy is expressed in terms of the velocity instead of the wavenumber. 

Comparing the two sets of curves, we see that the system does, indeed, retrieve a more linear behaviour in the high energy regime where the effect of kinetic energy is greater than that of the non-linear energy. While the curves do not quantitatively align in the range displayed, they at least share a qualitative agreement. At lower energies, where we see bunching/convergence of the red iso-curves which illustrates disallowed states, the transmission behavior departs from being comparable to the linear system and becomes truly non-linear.

In ~\figreft{fig:pscan}{(b)} we display curves of transmission as a function of velocity for a range of values of $q$. The shaded (red) region is the region of $T_+,v$ combinations disallowed under inequality~\eqnreft{eqn:VSCOND}. In the high kinetic energy regime these curves increase monotonically, but at low kinetic energies this ceases to be true~\cite{Gertjerenken_etal_pra_2012,POLO_etal_PRA_2013,Molmer_arxiv_2012,WangEtAl2012}. We see that here (in the absence of an axial harmonic trap) the disallowed region is quite strict, with no substantial violation of ~\eqnreft{eqn:VSCOND}. Indeed,~\eqnreft{eqn:VSCOND} is generally found to be more strict than the numerical result, as seen by the empty gaps between the disallowed region and the transmission curves.

In ~\figreft{fig:pscan}{(c)}, the last part that pertains to the axially untrapped case, we display curves of transmission as a function of barrier strength for a range of collisional/initial velocities. The labelled (red) curve, for which $v=0.25$, shows the bound below which there is never enough kinetic energy to access all splitting events. We see that all curves $T_+(q;v)$ are discontinuous for $v\le0.25$, although the discontinuous region is narrower for higher $v$, and is instantaneous for the $v=0.25$ case. 

~\figreftfull{fig:pscan}{(d--f)} are the harmonically trapped counterparts of the figures described above, as we described at the beginning of this subsection. The behaviour is broadly the same, however there are some specific qualitative and quantitative differences. 

In terms of qualitative differences, we see in~\figssreft{fig:pscan}{(e)}{(f)} that there exists a class of solution which appears to access disallowed outcomes, shown by points lying within the shaded (red) region of the plot. Upon closer inspection we determined these outcomes to be bound state solutions~\cite{POLO_etal_PRA_2013}. The energetic arguments leading to~\eqnreft{eqn:VSCOND} suppose that the solitons are, after splitting, well separated. If this is not the case then we can access a bound state solution. In this event, the kinetic energy shortfall (the deficit of energy required to fully split the soliton) is made up for by the bound-state interaction energy which is gained from the overlap, and attraction, between the resulting solitons. This effect can be greatly enhanced in the harmonically trapped system, where an insufficient kinetic energy after splitting means that the solitons cannot fully separate in the trap, necessitating a bound state.

Quantitatively we see that the value of $v_0$ (the velocity of the soliton at the bottom of the trap in the absence of a splitting potential, which we take to be the collisional velocity) must be slightly higher than its untrapped counterpart $v$ in order to access a continuous range of splitting outcomes. This is because the soliton begins to interact with the barrier slightly before it reaches the bottom of the trap at $x=0$, and so the collisional velocity is, in fact, slightly lower than $v_0$. This is shown by: the gap between the transmission curves and the disallowed region being wider in ~\figreft{fig:pscan}{(e)} than in~\figreft{fig:pscan}{(b)}; and the labelled (red) transmission curve in~\figreft{fig:pscan}{(f)} having a substantially wider discontinuous region than its counterpart in ~\figreft{fig:pscan}{(c)}, where the GPE limit of $N\to\infty$ is taken.

\subsection{Classical indicators of the transition to the quantum regime}
The behaviour we observe here, which describes an energy bound below which the possibility for splitting to occur is progressively curtailed, mirrors behaviour which leads to the generation of entangled states~\cite{Gertjerenken_prl_2013} in the purely quantum mechanical treatment. Indeed, it has been shown that entangled states in the fully quantum mechanical imply the discontinuities we see here~\cite{Gertjerenken_etal_pra_2012}. There is also evidence for the reverse implication~\cite{Gertjerenken_prl_2013}, and so it is conceivable that these behaviours are equivalent to the extent that transmission discontinuities in the mean-field treatment delimit the regime where mesoscopic Bell states would exist in the fully quantum mechanical treatment, despite these states not being present in the GPE formalism.

\subsection{Analysis of the effect of quantum uncertainty\label{sec:quant_analysis}}
  \begin{figure}[t]
    \includegraphics[width=\columnwidth]{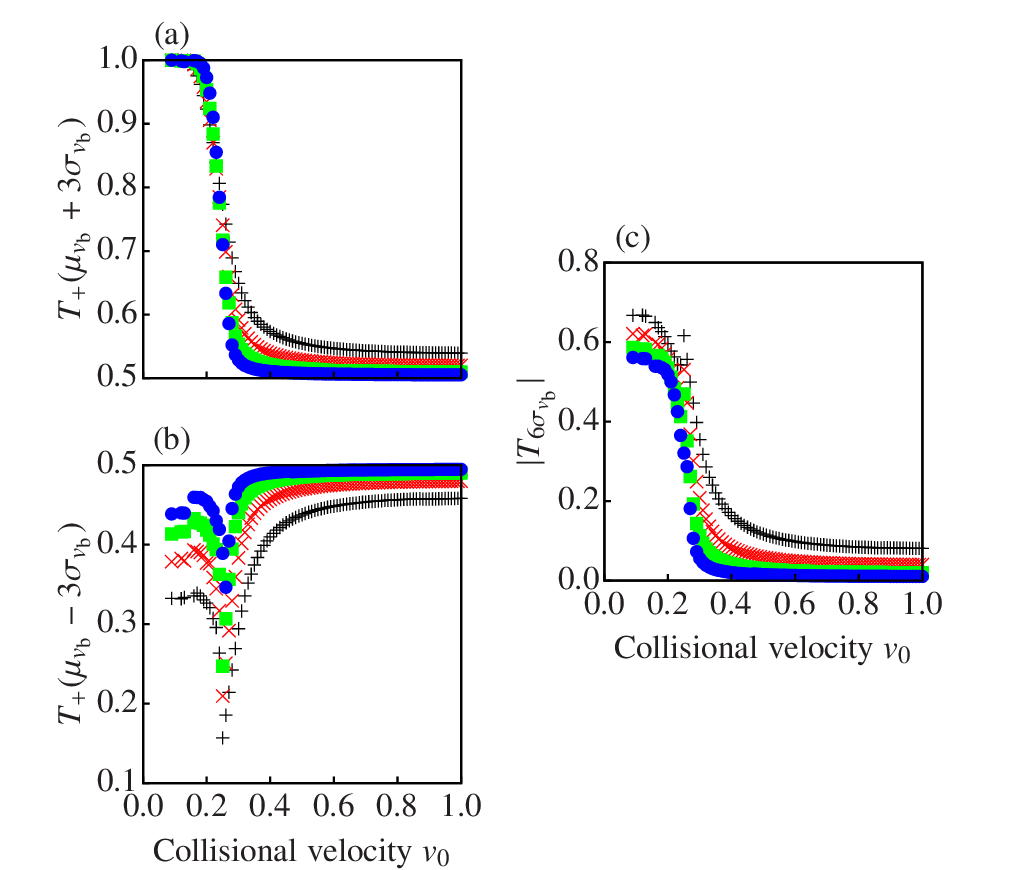}
    \caption{(Color online) Results of numerical integrations of the GPE illustrating the sensitivity of equal splitting to extreme quantum fluctuation for various particle numbers. The transmission after extreme positive (negative) energy quantum fluctuations are displayed in panel (a) (panel (b)). The number fluctuation measure $T_{6\svb}$~\eqnrefp{eqn:t_sen} is plotted in (c). For all plots we show $N=16$ ($\color{black}{+}$), $32$ ($\color{red}{\times}$), $64$ ($\color{green}{\blacksquare}$) and $128$ ($\color{blue}{\bullet}$).}
    \label{fig:num_unc_s}
  \end{figure}

We now address this high- to low-energy transitional regime by considering how quantum uncertainty impacts the dynamics of the system. The transmission through the barrier is determined by the velocity of the soliton at the point of collision. In the harmonically trapped system, fluctuations in the initial COM position and momentum will affect this velocity and so affect the transmission. We consider these uncertainties in the harmonically trapped system only, which presents a better defined situation than the untrapped, periodic regime when considering quantum fluctuations of the COM. In order to delimit a regime where the position/momentum uncertainty of the soliton affects the outcome of a splitting event, we must develop a formalism which allows us to introduce this uncertainty into our system.

First we consider a full many body treatment of our 1D $N$-particle system. We can write the first quantized form of the Hamiltonian as~\cite{Liniger_1963}
  \begin{equation}
    \hat{H}(\vec{x})=\sum_{k=1}^{N}\left(-\frac{\hbar^2}{2m}\frac{\pa^2}{\pa x_k^2}+\frac{m\Tom^2 x_k^2}{2}\right)-g\sum_{k=2}^{N}\sum_{j=1}^{k-1}\delta(x_k-x_j).
  \end{equation}
In this notation, $\vec{x}$ denotes the vector of the positions of all $N$ particles, $\{ x_1,x_2,..,x_N\}$, and all quantities are expressed in their fully dimensional form.

Moving to Jacobi coordinates we can show that the center of mass (COM) dynamics and the internal degrees of freedom separate~\cite{Holdaway_etal_2012} by expressing the Hamiltonian as $H=\HC + \HR$, where
  \begin{equation}
    \HC(\xc)=-\frac{\hbar^2}{2Nm}\frac{\pa^2}{\pa \xc^2}+\frac{Nm\Tom^2\xc^2}{2}
  \end{equation}
is simply the single particle Hamiltonian for a particle of mass $Nm$ at position $\xc$ -- the COM coordinate. $\HR$ describes the residual internal dynamics.

The dimensional wave function for the COM, $\PC$, is then given by 
  \begin{equation}
    \PC(\xc)=\left(\frac{1}{\tilde\Sx\sqrt{2\pi}}\right)^{1/2}\exp{\left(-\frac{\xc^2}{4\tilde\Sx^2}\right)}.
  \end{equation}
which is simply the 1D wave function of a single particle of mass $mN$ in an axial harmonic trap of frequency $\Tom$ normalized to 1. We can interpret $|\PC|^2$ as the probability density function for the normally distributed random variable $\xc$ such that the expected value is $\inp{\xc}=0$ and the variance (or the position uncertainty of our soliton) is given by $\inp{\xc^2}=\tilde\Sx^2=2mN\Tom/\hbar$. 

For our purposes, it is better to consider velocity uncertainty than it is to consider momentum uncertainty. Regardless, we must express our COM wave function in momentum space to obtain the momentum/velocity uncertainty. We now use standard result for the Fourier transform of a Gaussian, giving us the Fourier space wave function
  \begin{equation}
    \FC(\kc)=\left(\frac{1}{\tilde\Sk\sqrt{2\pi}}\right)^{1/2}\exp{\left(-\frac{\kc^2}{4\tilde\Sk^2}\right)}.
  \end{equation}
where the wavenumber variance is $\inp{\kc^2}=\tilde\Sk^2=1/4\tilde\Sx^2=mN\Tom/2\hbar$. We can now determine the momentum uncertainty ($\hbar\Sk$) and so the velocity uncertainty $\tilde\Sv=(\hbar/mN)\tilde\Sk$.

Rescaling the position and velocity uncertainties into dimensionless quantities, we now have 
  \begin{equation}
  \begin{split}
    \Sx&=(1/2N\xom)^{1/2},\\
    \Sv&=(\xom/2N)^{1/2}.
    \label{eqn:unc}
  \end{split}
  \end{equation}
These uncertainties are consistent with the GPE formalism in that as $N\to\infty$ they both disappear. In this limit, the full wave function $\psi$ gives the actual density profile, rather than a probability density function. As such, the COM and velocity distribution can be exactly determined. 

We now consider this system with an initial condition described by a ground state soliton at position $x_0$. If we consider a single observation of the quantum system, we see that the soliton's initial position and velocity are given by $x_0+\xf$ and $\vf$, where $\xf$ and $\vf$ denote the quantum fluctuations and are, therefore, normally distributed random variables with mean $0$ and standard deviations $\Sx$ and $\Sv$ respectively. By classically evolving these initial conditions (according to~\eqnreft{eqn:1DGPE}) we can apply previous results to state that the final transmission will depend on the fluctuating collisional velocity $\vb$, where $\vb=[\xom^2\left(x_0+\xf\right)^2+\vf^2]^{1/2}$.

By re-writing this velocity as $\vb=[\fomega^2+\vf^2]^{1/2}$, where $\fomega=\xom(x_0+\xf)$, we can see that $\vb$ is essentially the length of a vector comprised of two normally distributed random variables: $\fomega\sim\mathrm{N}(\xom x_0, \Sv^2)$ and $\vf\sim\mathrm{N}(0,\Sv^2)$. Note that both variables are Gaussian and have the same variance. As such, we can treat the collisional velocity $\vb$ as a Rician distributed random variable $\vb\sim\mathrm{R}(\xom x_0,\Sv)$, and so is described (in terms of the Laguerre polynomials of order $1/2$, $\mathrm{L}_{1/2}$) by mean and variance $\mvb, \svb$ defined as
\begin{align}
  \mvb=\E{\vb}=\Sv\sqrt{\frac{\pi}{2}}\mathrm{L}_{1/2}\left(\frac{-(\xom x_0)^2}{2\Sv^2}\right),\\
  \svb^2=\V{\vb}=2\Sv^2+(\xom x_0)^2-\mvb^2.
\end{align}

\subsection{Numerical analysis of the effects of quantum uncertainties\label{sec:quant_num}}
  
\begin{figure*}[t]
    \includegraphics[width=\textwidth]{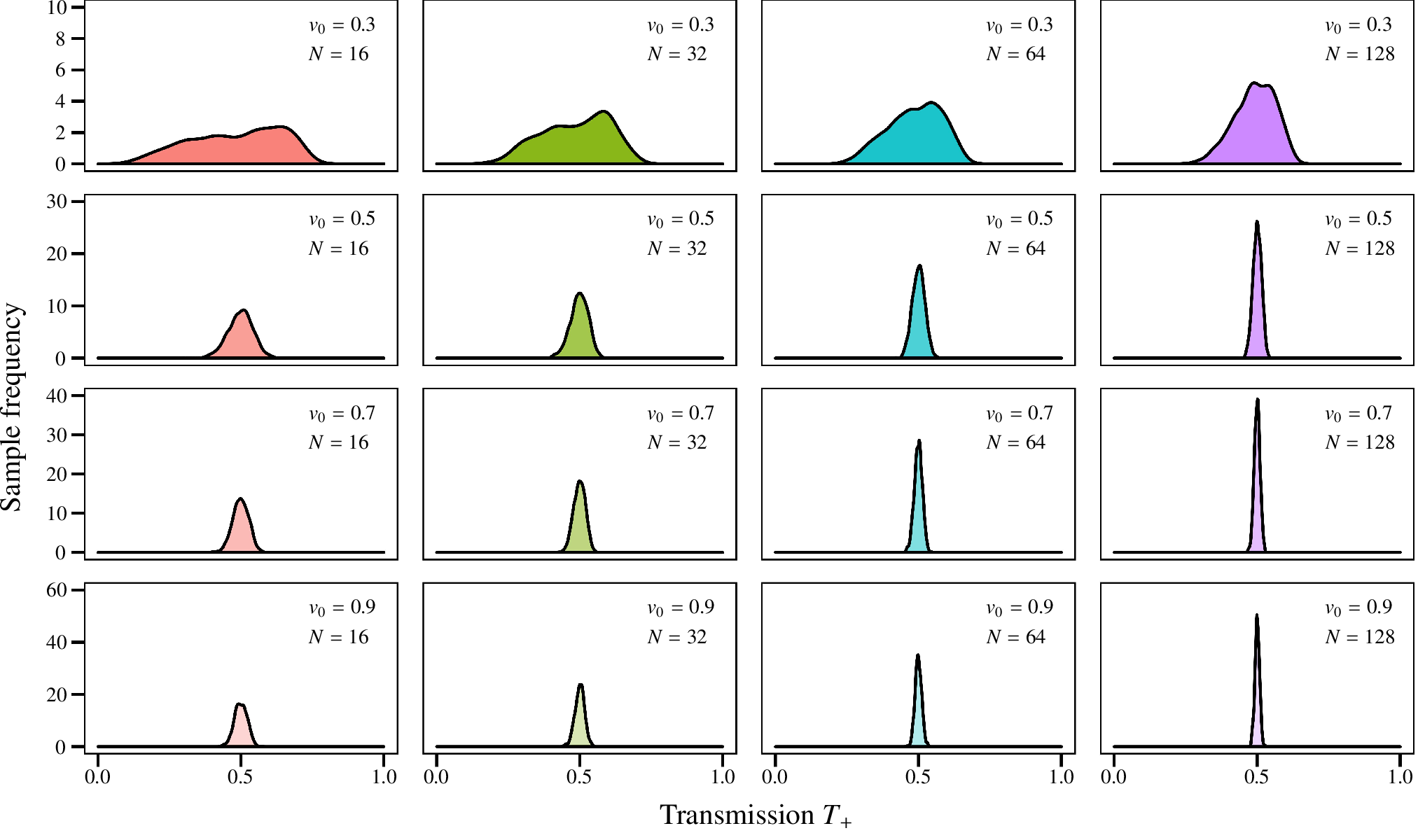}
    \caption{(Color online) Distributions of the transmission $T_+$ obtained from Monte Carlo simulations. Here we show results for a range of trap frequencies and particle numbers, giving a range of uncertainties in the initial COM position and momentum. In the range explored, we see that the effects of varying the trap frequency (and so kinetic energy) dominate the dynamics, with narrow Gaussians at high energy, but a bimodal structure arising at low energy when energetically disallowed states arise.}
    \label{fig:pdf_s}
  \end{figure*}

\subsubsection{Overview of the method\label{sec:quant_num_over}}

We now wish to characterise the effect of COM and collision velocity uncertainties on the soliton's transmission through the barrier after being accelerated by the harmonic trap ($T_+$). To determine the effect of these quantum fluctuations we perform a Monte Carlo analysis, where we numerically evolve the GPE~\eqnrefp{eqn:1DGPE} with fluctuations in the initial COM position and momentum. This procedure uses the COM truncated Wigner approximation (TWA), as used in Ref.~\cite{Gertjerenken_prl_2013} to describe the behavior of mesoscopic quantum superpositions. The COM TWA was shown to agree well with the effective potential approach of Ref.~\cite{weiss_castin_prl_2009}, demonstrating the validity of this method for describing quantum fluctuations in bright soliton systems. Note the related work investigating bright solitons using the TWA in Refs.~\cite{dabrowska_wuster_etal_njp_2009,martin_ruostekoski_njp_2010}.

To characterise the effects of quantum fluctuations, we performed numerical calculations of soliton splitting for varying particle numbers and trap frequencies. We perform these calculations over the same range of velocities as that explored in~\secreft{sec:class_num}, allowing for comparisons over the same energetic regime.

Given that this is the velocity range of interest we must select a range of values for the particle number $N$ such that the relevant uncertainties~\eqnrefp{eqn:unc} generate fluctuations which are significant relative to the grid spacing in the numerical algorithm. With $4096$ spatial grid points over a $-20\pi<x<20\pi$ domain we have a grid spacing $\Delta x\approx0.031$. If we now require that $\Sx/\Delta x>10$ (giving twenty grid points within one standard deviation of the spatial mean), we are limited to $N\lesssim166$. We will distribute $N$ logarithmically over this range (taking powers of $2$) and so we consider $N={16,32,64,128}$. 

It should be noted that this limit on $N$ was determined with $v_0=1$, and so in general there are significantly more than twenty grid points within one standard deviation of the mean. For example, with $N=16$ and $v_0=0.1$ there are over two hundred grid points within one standard deviation of the mean.

In both sections, for each value of $v_0$ a value of the barrier strength $q$ was selected such that the soliton would be split equally in the absence of quantum fluctuations on the initial condition. The barrier's width was $\sigma_b=0.2$ for all runs. 

\subsubsection{Transmission sensitivity to quantum fluctuations\label{sec:trans_sens}}

  \begin{figure}[t]
    \includegraphics[width=\columnwidth]{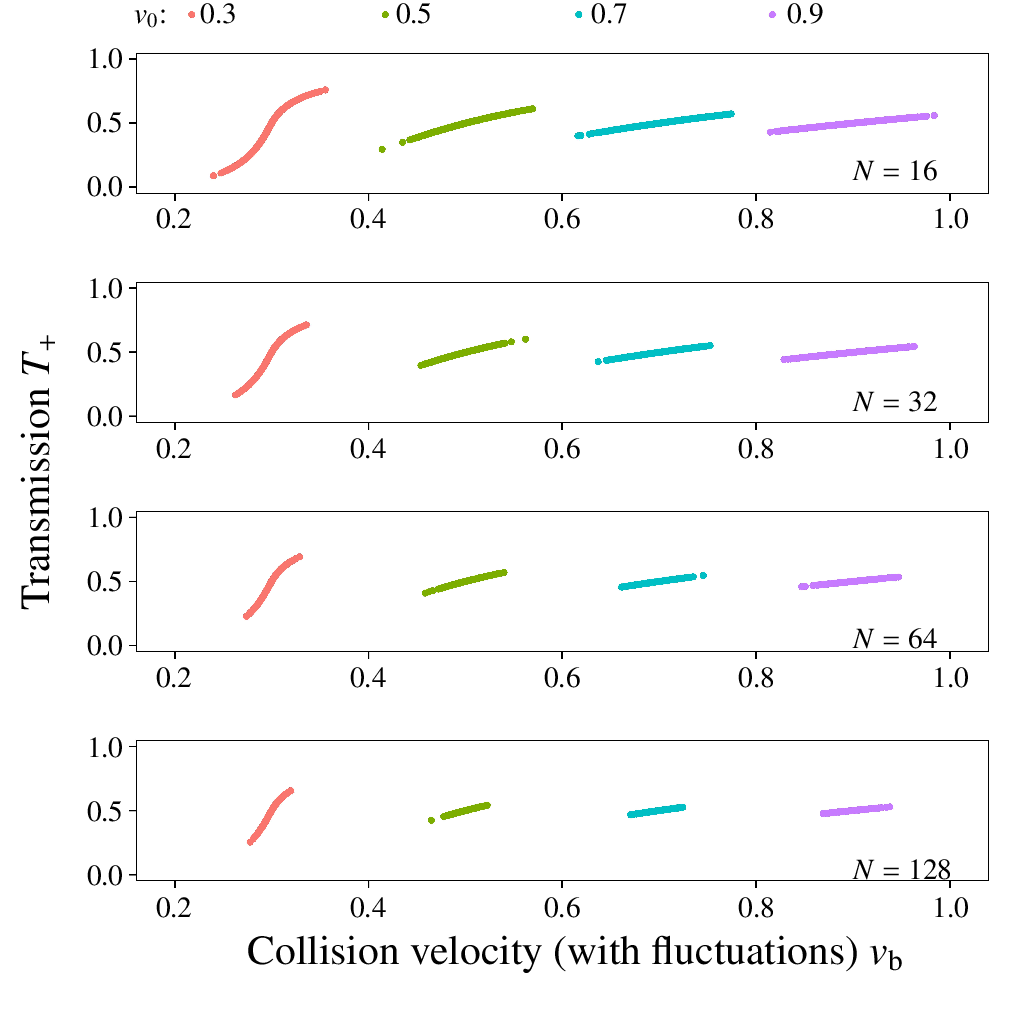}
    \caption{(Color online) Results of Monte Carlo simulations. Here we show the dependence of transmission on $T_+$ on the collision velocity ($\vb$) after quantum position/momentum fluctuations have been added to a base collision velocity ($v_0$). For each $v_0$ the barrier strength was set to ensure equal splitting in the limit of zero fluctuations. We see that in the low energy regimes the transmission can be very sensitive to quantum fluctuations.}
    \label{fig:tdist}
  \end{figure}

  \begin{figure}[t]
    \includegraphics[width=\columnwidth]{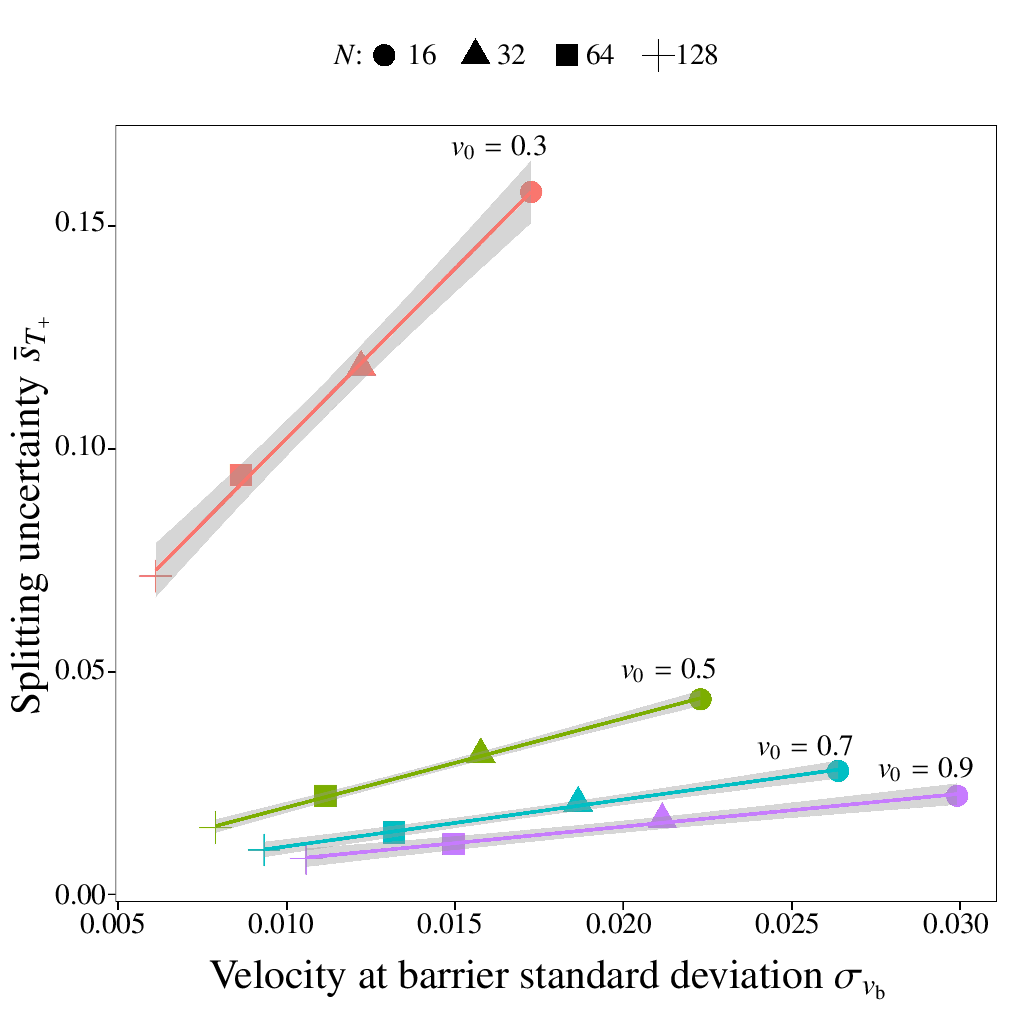}
    \caption{(Color online) Results of Monte Carlo simulations. Here, the standard deviations associated with the final transmission distributions depicted in~\figreft{fig:pdf_s}. We see a weak linear dependence on the sample velocity uncertainty $\bar s_{v_\mathrm{b}}$ for high $v_0$, which becomes stronger, but less linear, as we reduce the energy. This can be seen by the widening (shaded) $95\%$ confidence intervals of the linear fits.}
    \label{fig:sigmas_s}
  \end{figure}

We first characterise the sensitivity of the equal splitting case to extreme quantum fluctuations over a continuous range of $v_0$. For $v_0$ in the range $0<v_0 \le1$ the barrier strength $q$ was found such that $T_{+}(v_0)=1/2$. The simulation was then run twice more, replacing the initial position $x_0$ with $x_\pm=(\mvb\pm3\svb)/\xom$~\figrefp{fig:num_unc_s}{(a,b)}. This selection achieves collisional velocities at the barrier of $\mvb\pm3\svb$. The transmissions associated with these initial conditions [$T_+(\mvb\pm3\svb)$] illustrate the effects of extreme quantum fluctuations. These velocities represent extreme cases of quantum uncertainties adding/removing energy from the system, and so the $+/-$ cases correspond to extreme positive/negative energy quantum fluctuations in the system, and will be referred to as such hereafter.

We have also constructed the number fluctuation measure 
\begin{equation}
   T_{6\svb}=|T_+(\mvb+3\svb)-T_+({\mvb-3\svb})|.
   \label{eqn:t_sen} 
\end{equation}
This measure takes values between $0$ and $1$, with $0$ indicating absolute insensitivity to fluctuation and $1$ indicating a complete population shift resulting from extreme fluctuations in the initial COM position and momentum. 

~\figreftfull{fig:num_unc_s}{(a)} shows that $T_+(\mvb+3\svb)$ behaves as we might expect. As the collisional kinetic energy of the system decreases (shown by decreasing $v_0$), we see that extreme fluctuations in the initial COM position and momentum cause a deviation from from equal splitting. At first, when $v_0$ is relatively high ($v_0\gtrsim0.5$), the deviation of $T_+$ from $0.5$ is weakly dependent on $v_0$. Then, as $v_0$ approaches $0.25$ the effect of disallowed states becomes dominant. In this regime we see that extreme positive energy quantum fluctuations rapidly enhance transmission.

The effects of extreme negative energy quantum fluctuations, quantified by $T_+(\mvb-3\svb)$, are slightly more complicated. The careful selection of $q$ makes the bound states (as described in ~\secreft{sec:class_analysis} and observed in ~\secreft{sec:class_num}) a notable factor. This can be seen by the more complex structure of the data displayed in~\figreft{fig:num_unc_s}{(b)}. At the high energy end of the velocity range we see the same weak deviation of $T_+$ from $0.5$ as that described above for extreme positive energy fluctuations. However, where we might expect disallowed states to enhance reflection (namely $v\lesssim0.25$), we see a revival in the transmission. This is a result of a bound state confining the wave function to the region around the barrier at the bottom of the trap, resulting in a $T_+$ failing to tend to $0$. This effect is consistent with the reduced kinetic energy being insufficient to split the soliton in the low velocity regime.

Finally, in ~\figreft{fig:num_unc_s}{(c)} we see that $T_{6\svb}$ does generally increase as $v_0$ decreases, showing that number fluctuations become very important at low kinetic energies as a result of energetically disallowed states enhancing transmission/reflection. However, as a result of the previously discussed impact of bound states, $T_{6\svb}$ does not vary smoothly between $0$ and $1$. This effect could be treated as an artefact and removed by only taking the post-splitting positive domain integral ($T_+$) far from the barrier, thus excluding bound states. This would give a continuous, smooth range between $0$ and $1$, but would obscure the effect of bound states.

\subsubsection{Monte Carlo analysis of transmission with quantum fluctuations\label{sec:montecarlo_split}}

  \begin{figure}[t]
    \includegraphics[width=\columnwidth]{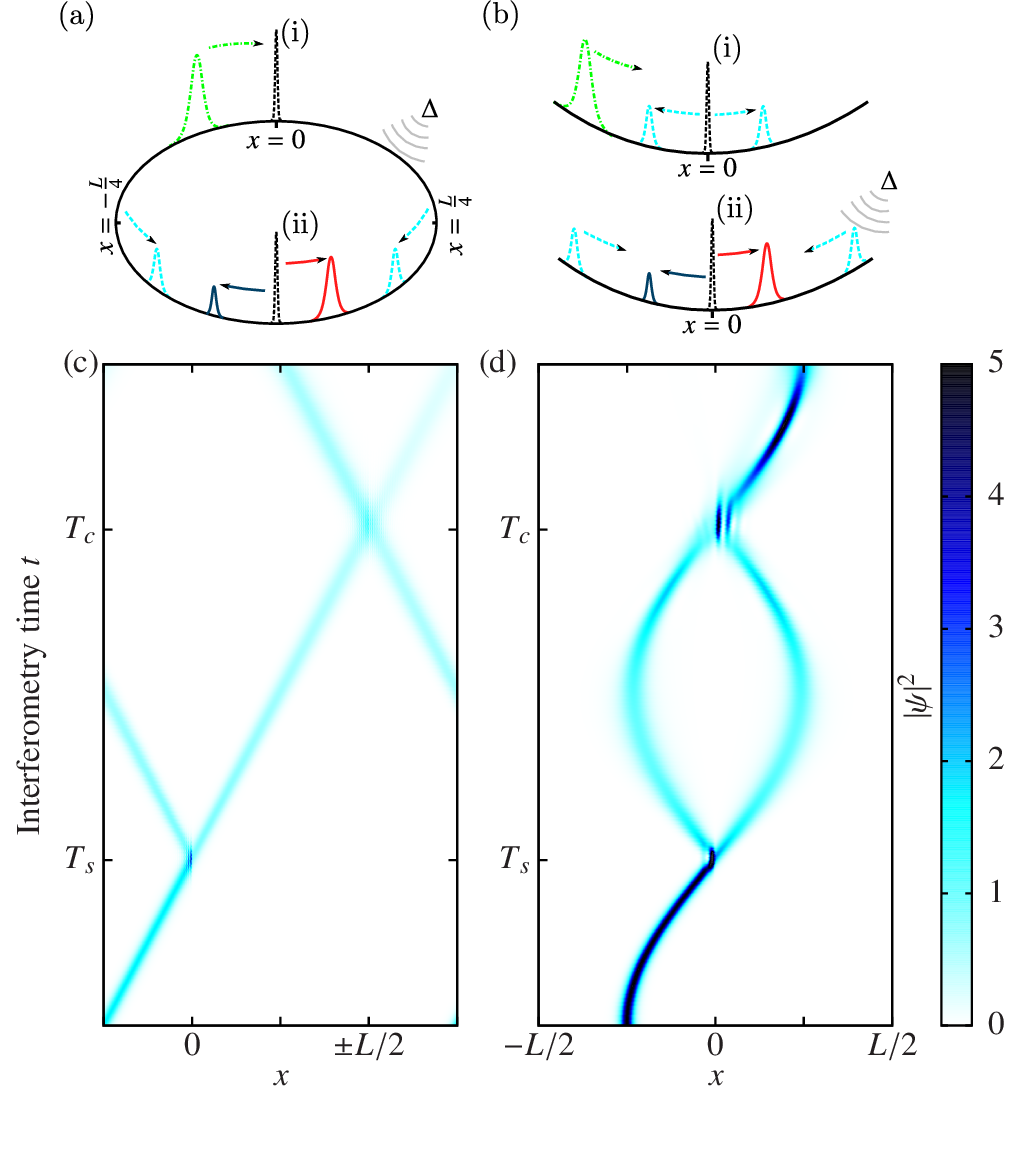}
    \caption{(Color online) (a) Diagram of a Mach-Zehnder interferometer utilising a periodic confinement with two antipodal barriers. An example of the time evolution of the density for this configuration is displayed in (c). (b) Diagram of a Mach--Zehnder interferometer utilising harmonic confinement and a single splitting barrier. Again, an example of the time evolution for such a configuration is displayed in (d).}
    \label{fig:plan}
  \end{figure}

In order to characterise the distribution of the transmission $T_+$ after factoring in quantum uncertainty in the initial condition we performed a selection of Monte Carlo simulations. These simulations allow us to develop a broader qualitative understanding of the effects of quantum uncertainty. Here we have selected the same values of the particle number $N$ as used previously and consider velocities $v_0={0.3,0.5,0.7,0.9}$. We present the results of $1000$ Monte Carlo simulations for each $v_0,N$ pair. 

\figreftfull{fig:pdf_s}~displays the different distributions of the transmission $T_+$ which arise from varying the energetic regime and particle number. In the bottom row we see that for high $v_0$ the distribution is a narrow Gaussian for all displayed $N$. Reducing $v_0$ for a given $N$ (reading up the column) causes the standard deviations of the Gaussians to broaden. For $v_0=0.3$ (the top row of~\figreft{fig:pdf_s}) a bimodal distribution appears, again illustrating that the equal splitting case is less easily accessed. This behaviour is evident for all $N$. Reading across the rows (varying $N$ while keeping $v_0$ constant) shows that increacing $N$ simply reduces the width of the transmission distribution. This illustrates that the $N$ dependence is secondary to the $v_0$ dependence in the range explored here. This is evident in that there is still significant broadening of the transmission distribution at low $v_0$ even for the highest values of $N$. We might expect this to be the case, given that the range of $N$ explored here is, in experemental terms, very low.

We can see the functional dependence of transmission on $\vb$ ($T_+(\vb)$) in~\figreft{fig:tdist}. We see that in the higher energetic regime ($v_0>0.5$) the transmission has a weak approximately linear dependence on the velocity. The relatively small gradient of this dependence indicates that the transmission is less sensitive to the fluctuations. For the $v_0=0.3$ data we see that the dependence becomes very sensitive to small fluctuations around $\vb=0.3$, the equal splitting case. This confirms that proximity to the energetically disallowed state can cause large variations in transmission when quantum fluctuations are considered. Increasing $N$ has the effect of narrowing the distributions of the fluctuations, and so these fluctuations can affect the transmission less dramatically, even when close to the energetically disallowed state. It should be noted that the points in~\figreft{fig:tdist}{} lie along curves with structure analogous to those depicted in~\figreft{fig:pscan}{(e)}. 

We can quantify the relationship between the initial quantum uncertainties (via $\svb$) and the resulting transmission uncertainty $\sigma_{T_+}$ by making a maximum likelihood estimate $\bar s_{T_+}$ based on our data. We assume that the data follows a truncated Gaussian distribution on the interval $[0,1]$. The results of these estimates are shown in~\figreft{fig:sigmas_s}. We see that $\bar s_{T_+}$ has approximately linear correlations with $\sigma_{v_\mathrm{b}}$. This correlation becomes stronger, illustrated by the increased gradient of the linear fit, as we reduce $v_0$. The grey shaded areas indicate a $95\%$ confidence interval for the least squares linear regression. The confidence interval associated with $v_0=0.3$ is widest, indicating a less linear relation between $\svb$ and $\bar s_{T_+}$ in the low energy regime.

  \begin{figure}[t]
    \includegraphics[width=\columnwidth]{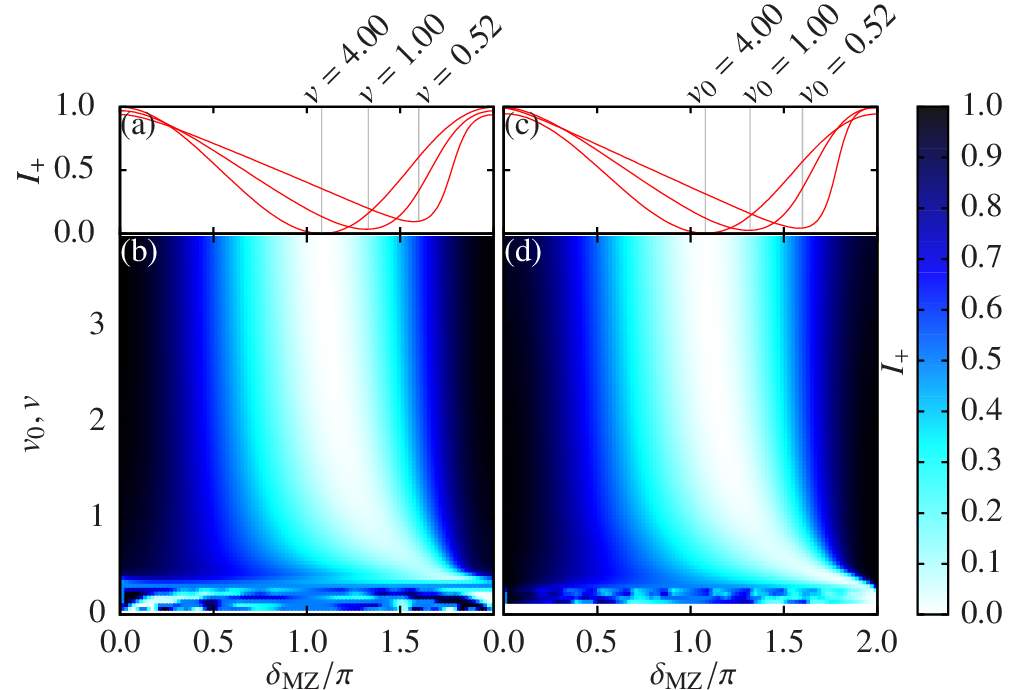}
    \caption{(Color online) Numerically calculated transmission rates after the second collision, $I_+$, for two Mach-Zehnder interferometry geometries. Color-maps for the (b) toroidal Mach-Zehnder and (d) harmonic Mach--Zehnder cases show the full parameter space. (a) and (c) show specific curves of constant $v$, $v_0$ for the same respective scenarios and highlight the transition from the high energy sinusoidal dependence regime to the lower energy quasi-linear dependence regime.}
    \label{fig:int}
  \end{figure}

\subsection{Split induced phase shift\label{sec:split_shift}}
In order to construct an analysis of soliton interferometry there is another aspect of soliton splitting which we must address. The act of splitting the soliton (which gives us two coherent matter waves to interfere) causes a phase difference to arise between the solitons. This is similar to the case of classical optics. A classical analysis of electromagnetic fields at interfaces between media, yielding the Fresnel equations~\citep{hecht_4th_ed}, shows us that when light passes into a medium with a higher refractive index the reflected part is phase shifted by $\pi$ with respect to the transmitted part. This effect is particularly relevant in the case of optical interferometers where a beam of light is split by a beam splitter. In the case of soliton splitting the principle is similar, and the barrier (here acting as our beam splitter) imparts a phase difference between the two residual solitons. In contrast to the optical case, the transmitted soliton is $\pi/2$ phase shifted with respect to the reflected soliton. In other words, the phase difference has half the magnitude and opposite sign. This difference between the two cases is understandable, as the two are very different physical systems and so are governed by very different sets of equations. The systems are analogous but, of course, not identical. We now present a derivation of this phase shift. 

It has been rigorously analytically shown~\citep{holmer_etal_cmp_2007} that, in the high kinetic energy limit (high soliton velocity $v$) of the 1D untrapped system, when a soliton is split at a $\delta$-function barrier the phases imparted to the solitons by the split are
 
  \begin{align}\label{phases1}
    \begin{split}
    \vartheta_{T}&=\left[1-A_{T}^2\right]\left|\frac{x_0}{2v}\right|+\arg\left(t_q(v)\right)+\vartheta_0\left(|t_q(+v)|\right),\\
    \vartheta_{R}&=\left[1-A_{R}^2\right]\left|\frac{x_0}{2v}\right|+\arg\left(r_q(v)\right)+\vartheta_0\left(|r_q(-v)|\right),
    \end{split}
  \end{align}
where $\vartheta_{R,T}$ are the reflected, transmitted soliton phases, and $A_{R,T}$ are the reflected, transmitted soliton amplitudes. Quantities $r_q(v)$ and $t_q(v)$ are the transmission and reflection rates of a $\delta$-function in the linear regime, given by
  \begin{equation}
      t_q(v)=\frac{\eye v}{\eye v -q} \qquad\mbox{and}\qquad
      r_q(v)=\frac{q}{\eye v -q}.
      \label{lin_trans}
  \end{equation}
If the barrier strength and initial velocity ($q$ and $v$) are selected to be equal ($q=v$), such that $|r_q(v)|=|t_q(v)|$ and (as a result) $A_R=A_T$ then the soliton is split equally into two secondary solitons of equal amplitude. This is desirable because later we will wish to collide the resulting solitons at a barrier, and if these solitons are of similar size then the interference between them is more pronounced. It is also true that a size difference causes secondary nonlinear phase shifts to arise during the collision, which is undesirable.

Making this selection, such that the soliton is equally split, and substituting appropriate values of $q$, $v$, $A_{R,T}$, $|r_q(v)|$ and $|t_q(v)|$ into~\eqnreft{phases1} we see that the relative phase between the solitons reduces to
  \begin{align}
    \vartheta_T-\vartheta_R&=\arg(t_q(v))-\arg(r_q(v))\notag\\
                       &=\pi/2.
    \label{eqn:split_shift}
  \end{align}

A broader discussion of the effect of a finite width barrier on the phase shift accumulated during splitting is, again, available in Ref.~\cite{POLO_etal_PRA_2013}.  We will use the $\pi/2$ figure as an estimate of the phase difference accumulated by splitting on a Gaussian barrier, as justified in~\cite{HELM_PRA_2012}, for the rest of the current work.
  
  \begin{figure}[t]
    \includegraphics[width=\columnwidth]{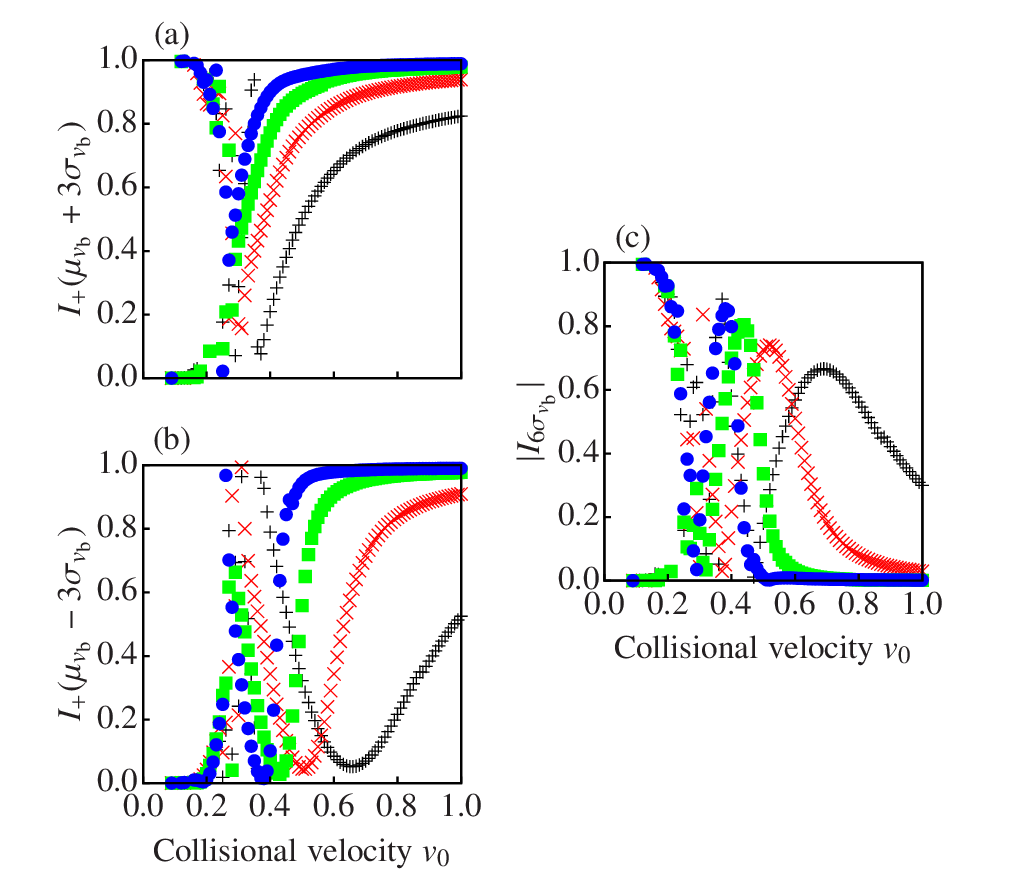}
    \caption{(Color online) Results of numerical integrations illustrating the sensitivity of interferometry to extreme quantum fluctuation for various particle numbers. The interferometry transmission after extreme positive (negative) energy quantum fluctuations are displayed in panel (a) (panel (b)). The number fluctuation measure $I_{6\svb}$ ~\eqnrefp{eqn:i_sen} is plotted in (c). For all plots we show $N=16$ ($\color{black}{+}$), $32$ ($\color{red}{\times}$), $64$ ($\color{green}{\blacksquare}$) and $128$ ($\color{blue}{\bullet}$).}
    \label{fig:num_unc_i}
  \end{figure}

\section{Soliton interferometry\label{sec:interferometry}}
\subsection{Analysis of soliton interferometry.}
We can use the above results regarding soliton interactions at narrow barriers to analyse and construct a soliton interferometer. Soliton interferometry is a three step process.

First we split a ground state soliton into two lesser solitons of equal size at a narrow potential barrier~\figrefp{fig:plan}{(a,b)(i)}. In the case of a $\delta$-function barrier, this split causes the transmitted soliton to gain a $\pi/2$ phase shift relative to the reflected soliton, as described in~\secreft{sec:split_shift}.

These solitons then accumulate a further relative phase difference $\Mdelta$. This phase difference is the quantity we wish to measure. In the current work we consider the case where this difference is gained by exposing one soliton to a phase shifting phenomenon.

In the third step the two solitons are made to collide at a narrow barrier~\figrefp{fig:plan}{(a,b)(ii)}. After this final barrier collision the wave-function integrals on either side of the barrier, 
  \begin{equation}
    I_{\pm}=\pm\int_0^{\pm\infty}|\psi(x)|^2\mathrm{d}x,
    \label{eqn:int_pm}
  \end{equation}
allow us to determine the magnitude of $\Mdelta$ ~\figrefp{fig:plan}{(a,b)(ii)}. Here $I_+$ is the positive domain population and $I_-$ is the negative domain population. We can determine the dependence of $I_\pm$ on $\Mdelta$ by recalling previous work by the authors~\cite{HELM_PRA_2012} in which it was shown that after two initially distinct solitons collide at a barrier, and had relative phase $\Delta$ before the collision, the populations in the negative and positive domains,
  \begin{equation}
    T_{\pm} = \pm\lim_{t\rightarrow \infty} \int_0^{\pm \infty} |\psi(x,t)|^2 \mathrm{d}x.
    \label{Idef}
  \end{equation}
are given by
  \begin{equation}
    T_\pm = \frac{1\pm\sin(\Delta+\epsilon)}{2},
    \label{epsilondef}
  \end{equation}
where
  \begin{equation} 
    \lim_{v\to\infty} \mbox{max}(\epsilon)=0.
  \end{equation}
Using this result we can see that taking the phase difference $\Delta$ to be the sum of the phase shift we wish to measure, $\Mdelta$, and the phase shift accumulated during the initial split, $\pi/2$ we obtain
  \begin{equation}
    I_\pm = \frac{1 \pm \cos(\Delta+\epsilon)}{2},
    \label{eqn:pop_pm}
  \end{equation}

The different types of soliton interferometry available are determined by the geometry of the potentials used to confine and split the BEC. Here we investigate two different geometries. The first is a toroidal trap giving a periodic geometry with two splitting potentials at antipodal points~~\secrefp{sec:tor_mac}~\figrefp{fig:plan}{(a,c)}. This geometry is somewhat challenging to create experimentally but provides the simplest framework in which to establish our analytical results. The second geometry uses a non-periodic geometry with a weak axial harmonic trap centered on a narrow splitting potential~\secrefp{sec:harm_mac}~\figrefp{fig:plan}{(b,d)}. This geometry makes is more experimentally viable, but questions of broken integrability require that we confirm the applicability of the results established above.

We will now present more expansive numerical analyses of these cases in order to determine whether our analytical results are confirmed numerically and also to determine the best energy regime in which to attempt soliton interferometry.

\subsection{Toroidal confinement Mach--Zehnder interferometry\label{sec:tor_mac}}
An often discussed trapping geometry is the periodic toroidal trap. The existence of experimental results utilising optical~\cite{Ramanathan_etal_prl_2011} and magnetic~\cite{Gupta_etal_prl_2005,Sauer_etal_2001} confinement methods coupled with theoretical investigations proving localised bright soliton states exist in mean-field/truncated Hamiltonian~\cite{Kavou_pra_2003}, 3--D GPE \cite{Parola_etal_pra_2005}, and coupled Gross--Pitaevskii Bogoliubov-de Gennes equations\cite{Sals_etal_pra_2007, Salas_etal_pra_2007} makes it worthwhile to consider extending our theory into this geometry. The toroidal geometry is beneficial in that it has no axial trapping, the presence of which breaks integrability and could, arguably, compromise our previous results~\footnote{Indeed, adding any potential breaks the integrability, but for narrow splitting barriers one can consider the system to be widely integrable with small regions where the solution behaves differently.}.

By treating~\eqnreft{eqn:1DGPE} as periodic over the domain $-L/2<x\le L/2$, such that $\psi(-L/2)=\psi(L/2)$, we obtain a suitable dynamics equation. We use the same initial condition~\eqnrefp{eqn:PODSECH} and initial offset, but set the trap frequency $\xom=0$ and directly vary the velocity $v$ by imprinting a phase on the initial condition.

Results of GPE simulations are shown in~\figssreft{fig:int}{(a)}{(b)}. We see that for very high velocities, $v\approx4$, the interference follows our prediction~\eqnrefp{eqn:pop_pm} closely, with very small skews arising from nonlinear effects during the final barrier collision, showing that $\epsilon\approx0$ in this regime.

As the velocity decreases, and we enter the transitional regime between high and low kinetic energy, $\epsilon$ increases and the skew becomes more prominent. As this happens the interference curve ceases to be sinusoidal and becomes approximately linear over some range, with $I_\pm\propto\mp\Mdelta$ up to some discontinuity. This discontinuity becomes narrower for higher $\epsilon$ and is situated at $2\pi$ for $v\approx0.3$. In this regime, however, we are drawing close to the regime where equal soliton splitting becomes disallowed. For $v\lesssim0.3$ the structure of the transmission becomes very complex, as the sensitivity of splitting to small changes in velocity becomes apparent. In this regime, soliton interferometry becomes impracticable.

  \begin{figure*}[t]
    \includegraphics[width=\textwidth]{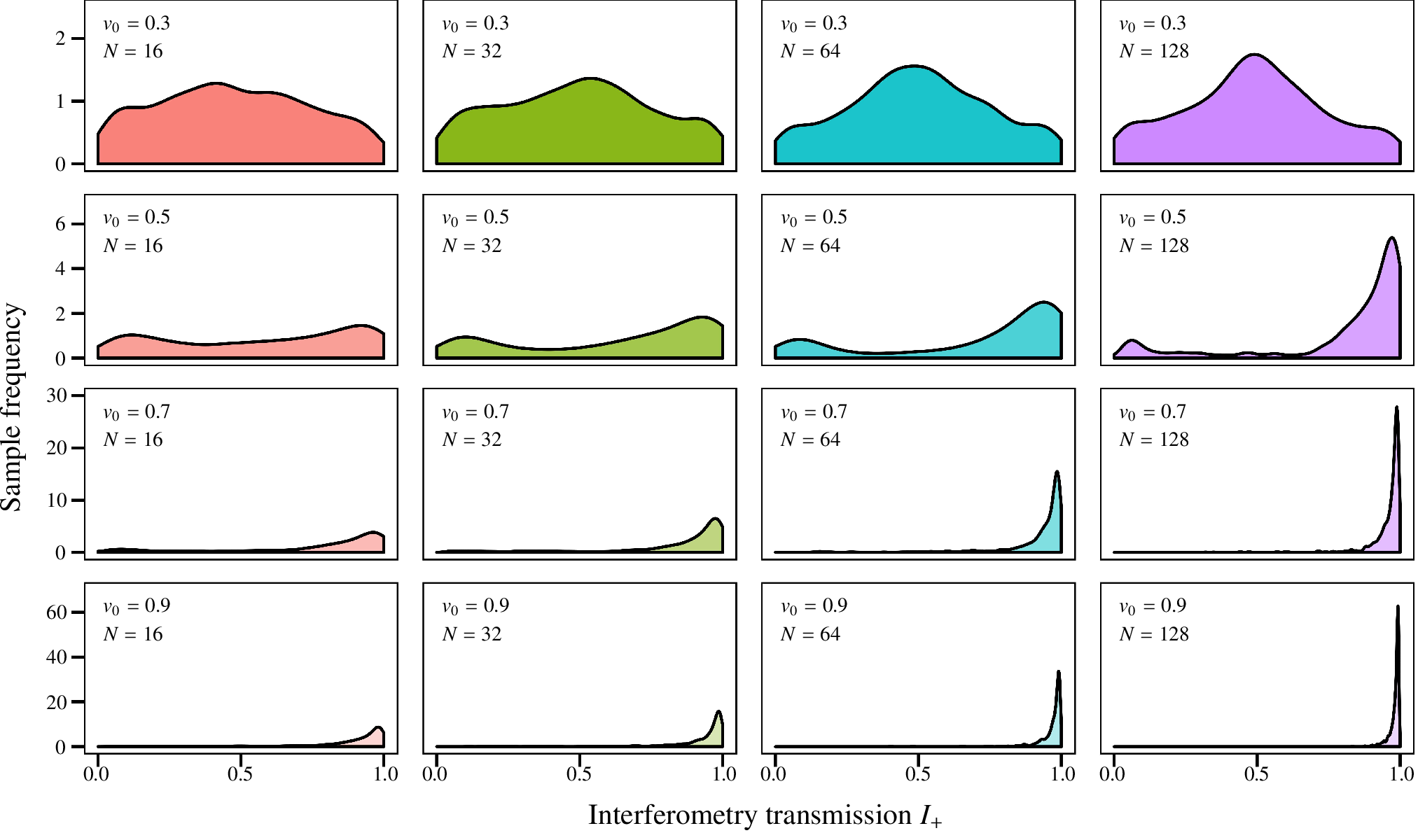}
    \caption{(Color online) Distributions of the interferometry transmission $I_+$ obtained from Monte Carlo simulations. Here we show results for a range of trap frequencies and particle numbers, giving a range of uncertainties in the initial COM position and momentum. In the range explored, we see that the effects of varying the trap frequency (and so kinetic energy) dominate the dynamics, with narrow Gaussians at high energy, but a uniform structure arising at low energy when interferometry becomes impracticable.}
    \label{fig:pdf_i}
  \end{figure*}

\subsection{Harmonic confinement Mach--Zehnder interferometry\label{sec:harm_mac}}

\subsubsection{Overview}
When considering trapping geometries for BEC experiments it is important to note that the addition of an axial harmonic trap globally breaks the integrability of the system, and so we can no longer say that we are studying true NLSE solitons in the mathematical sense. It is true, however, that the bright solitary waves supported by the system and confined in the harmonic trapping potential behave in a very soliton-like manner, staying robust to collisions and retaining their forms for long periods. Investigations utilising particle Hamiltonian models ~\cite{martin_etal_prl_2007} to describe the soliton motion agree well with GPE simulations, and so we can safely treat these bright solitary waves as solitons.

\subsubsection{Classical numerical analysis\label{sec:class_int}}

The results of fully classical numerical simulations are displayed in ~\figreft{fig:int}{(c)}{(d)}, obtained by evolving the initial condition described by~\eqnreft{eqn:PODSECH} according to~\eqnreft{eqn:1DGPE}. In this case, the initial velocity $v$ was set to zero while the soliton's velocity at the barrier, $v_0$, was set by varying the axial trap frequency $\xom$ (The dimensionless from of $\Tom$) and holding the initial offset $x_0$ at a constant value such that the soliton is initially well separated from the barrier. 

The results are comparable to those seen for the periodic Mach--Zehnder case~\secrefp{sec:tor_mac}, with good agreement with theory for high velocities, a linear dependence arising as we approach $v_0\approx0.3$ and finally complex structure arising in the low energy regime making interferometry impracticable\footnote{It should be noted that in the data set displayed in~\figreft{fig:int}{(d)} is incomplete. The solid white band at $v\approx0$ is a region where the system evolved too slowly to be numerically practical.}.

\subsubsection{Interferometry sensitivity to quantum fluctuations~\label{sec:sens_int}}

It was stated above that the linear relation between final domain population and phase shift might make interferometry more easily interpreted in the lower velocity regime. However, if we are to work in the regime we must consider the implications of the results outlined in~\secreft{sec:soliton_splitting}; namely the impacts of energetically disallowed states and quantum uncertainty in the initial condition.

We again characterise the system's sensitivity to extreme positive/negative energy fluctuations. As such, we construct the quantities $I_+(\mvb\pm3\svb)$ and 
\begin{equation}
   I_{6\svb}=|I_+(\mvb+3\svb)-I_+({\mvb-3\svb})|.
   \label{eqn:i_sen}
\end{equation}
These quantities are analogous to those used previously \secrefp{sec:trans_sens}, but are obtained by allowing the system to evolve through the entire process of interferometry, rather than just the initial splitting event. In this section and the next section [where we discuss results displayed in~\figssssreft{fig:num_unc_i}{fig:pdf_i}{fig:idist}{fig:sigmas_i}] we have considered the $\Delta=0$ case only in order to simplify our analysis.

~\figreftfull{fig:num_unc_i}~shows the results of these simulations. We see that for high $N$ and high $v_0$ the systems are reasonably insensitive to fluctuations. However, even in the high energy limit we see that as we decrease $N$ the interferometry transmissions significantly deviate from their asymptotic values. This sensitivity is high compared to that of the single splitting case, illustrating that the process of splitting (which occurs twice in interferometry) enhances the sensitivity of the classical system to initial fluctuations. The double enhancement in interferometry requires that we must be closer to the mean-field limit or suffer intolerable deviations from the classical behaviour.

As we decrease $v_0$ still further the previously discussed bound states and disallowed splitting events greatly complicate the dynamics of interferometry, making both the system and the results of our numerics difficult to interpret. This difficulty clearly shows that interferometry is impracticable in the low energy limit. 

\subsubsection{Monte Carlo analysis of interferometry with quantum fluctuations\label{sec:montecarlo_int}}
  
  \begin{figure}[t]
    \includegraphics[width=\columnwidth]{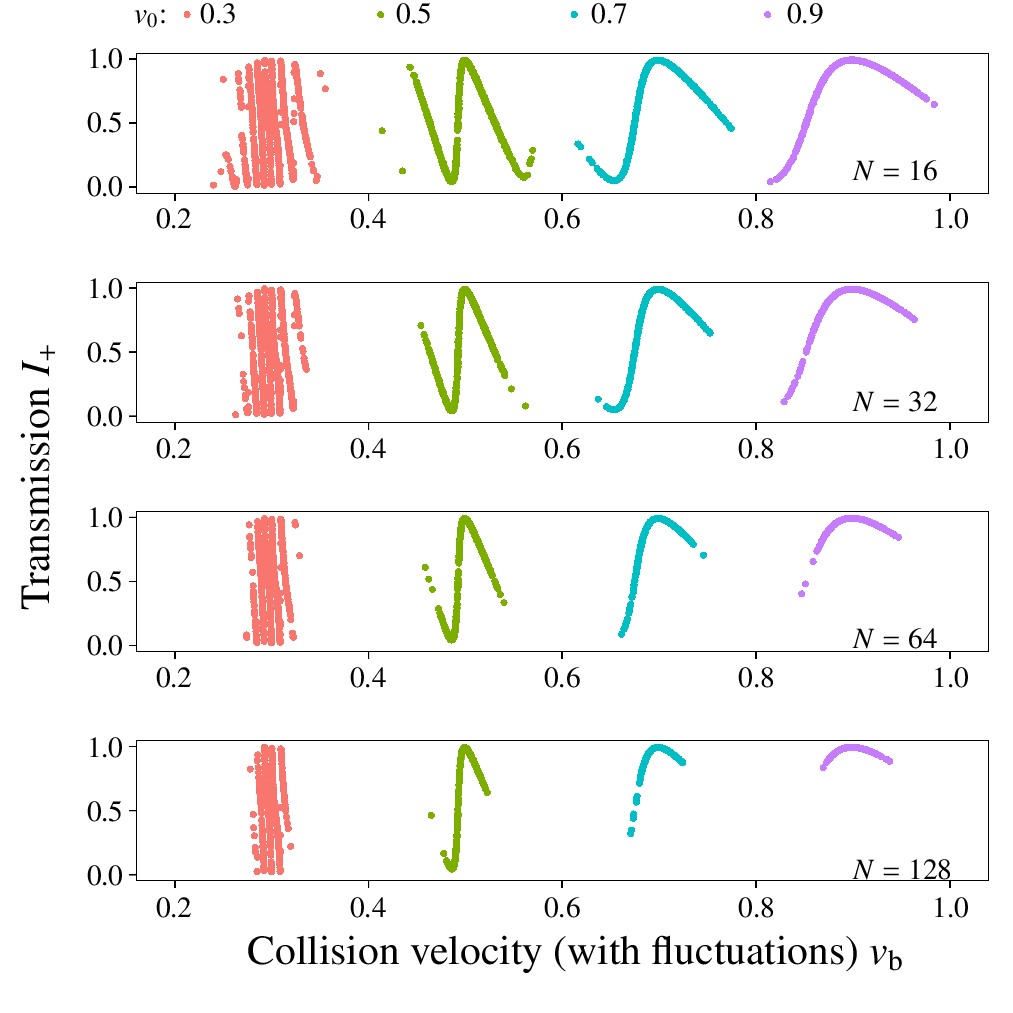}
    \caption{(Color online) Results of Monte Carlo simulations. Here we show the dependence of interferometry transmission on $I_+$ on the collision velocity ($\vb$) after quantum position/momentum fluctuations have been added to a base collision velocity ($v_0$). For each $v_0$ the barrier strength was set to ensure equal splitting in the limit of zero fluctuations. We see that in the low energy regimes the complex and velocity sensitive structure of the transmission renders interferometry unworkable.}
    \label{fig:idist}
  \end{figure}

We now present a Monte Carlo analysis of the effects of quantum uncertainties in the COM initial position and momentum. We explore the same parameter regime as in~\secreft{sec:montecarlo_split} and again present the results of $1000$ Monte Carlo simulations. The histograms in~\figreft{fig:pdf_i}, illustrating the distributions of the interferometry transmission $I_+$, show characteristics similar to those in~\figreft{fig:pdf_s}, but more pronounced. The distributions are approximately Gaussian at higher energies/particle numbers, but become more uniform at low energies ($v_0=0.3$), with a peak in the frequencies near $I_+=0.5$ arising from the presence of persistent bound states. This again indicates that interferometry is not viable in the low energy regime.

The transmission curves in ~\figreft{fig:idist}~have a much more complex structure than that exhibited in its counterpart~\figreft{fig:tdist}. At higher velocities, the points are clearly centered on the $I_+=1$ state, as we would expect, but as we lower the velocity the transmission becomes very sensitive to quantum fluctuations. This can be attributed to nonlinear phase shifts arising during the soliton collision at the barrier, compounded by a mis-match between the barrier strength and soliton velocity upon collision. Indeed, for the $v_0=0.3$ case these nonlinear phase shifts can cause $I_+$ to take literally any value between $0$ and $1$, and the quantum fluctuations cause $I_+$ to tune across this period multiple times. This, alone, precludes any possibility of soliton interferometry in this regime. It is also visible that, even for high energies, a particle number of less than $\approx130$ can cause increased sensitivity, and so we really must ensure that we are in the regime of high $N$. After these considerations have been taken into account, it should be possible to perform interferometry with a quasi-linear signal [similar to that associated with the $v_0=0.52$ curve in ~\figreft{fig:pscan}{(e)}] for values of $v_0\gtrsim 0.5$.

  \begin{figure}[t]
    \includegraphics[width=\columnwidth]{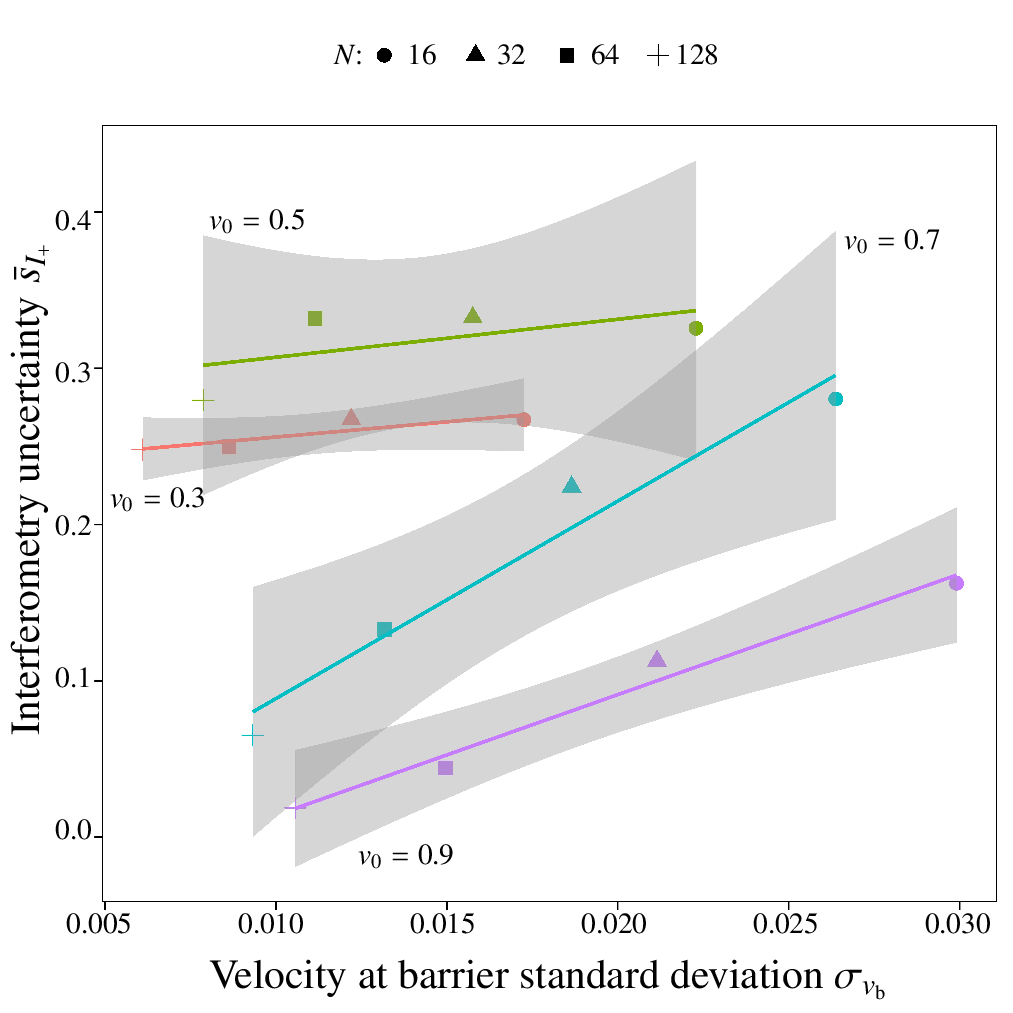}
    \caption{(Color online) Results of Monte Carlo simulations. Here, the standard deviation associated with the final interferometry distributions depicted in~\figreft{fig:pdf_i}. We see a strong, weakly linear dependence on $\bar s_{v_{\mathrm{b}}}$ for high $v_0$, which becomes stronger, but less linear, as we reduce the energy. The variance saturates when the distribution becomes effectively uniform.}
    \label{fig:sigmas_i}
  \end{figure}
  
Finally, we again calculated maximum likelihood estimates of the variance $\bar s_{I_+}$ of the transmission, which we again assumed to be distributed as a truncated Gaussian. The results of these calculations are displayed in~\figreft{fig:sigmas_i}.~At higher velocities, we see an approximately linear correlation between the transmission uncertainty and collisional velocity uncertainty standard deviation $\svb$. The gradient of the regression lines is much steeper than those in~\figreft{fig:sigmas_s}, showing the increased sensitivity of $I_+$ to quantum fluctuations. Again, the shaded regions show a $95\%$ confidence interval for the linear fit. For all velocities shown the confidence intervals are notably wider than their conterparts in~\figreft{fig:sigmas_s}, and so we can conclude that the dependence of $\bar s_{I_+}$ on $\svb$ is more complicated than in the soliton splitting case, as we would expect.  At lower velocities $\bar s_{I_+}$ saturates below $\sim 0.4$. This is a result of attempting to fit a Gaussian to a distribution which is, in effect, uniform. This becomes apparent when we consider that $\sim38\%$ of the probability mass of a Gaussian lies within a central period of width $\sigma$, and so applying a fitting algorithm to a uniform distribution will likely produce a standard deviation with a width encompassing $\sim38\%$ of the sample. In this case, that with is $\sim0.4$. This saturation is a strong indicator of a velocity/particle number regime in which interferometry is unworkable. 

\section{Conclusions}

We have shown analytic results describing soliton interferometry in the ideal classical case, specifically the case of a toroidal Mach--Zehnder configuration. We have extended these results to the harmonically trapped system, which is currently more experimentally relevant than the toroidal case~\cite{Marchant_etal_2013} and presents a better defined situation when considering quantum fluctuations of the COM. This has allowed us to investigate and delimit the energetic regimes in which quantum fluctuations in the initial COM position and momentum cause the classical dynamics to break down. 

This low energy regime failure of classical results is primarily caused by disallowed soliton splitting events, extremely discontinuous transmission curves, and bound states. These factors complicate the early evolution of the interferometric system and compromise the dynamics. As we approach the low energy regime quantum effects mix these phenomena into the dynamics of the system where classically they would be absent. This causes greatly enhanced sensitivity to quantum effects in both the splitting transmission and the interferometry transmission when close to the low energy regime. This sensitivity appears at marginally higher kinetic energies in the presence of harmonic trapping, but the difference is relatively slight for the weak trapping considered.

We conclude that whether or not the mean-field limit is truly achieved, soliton interferometry is not a viable process in the extremely, or even transitionally, low kinetic energy regime. However, for a suitibly high initial kinetic energy we see good results for particle numbers upwards of the low hundreds (beyond which our numerical algorithm struggles to resolve fluctuations, also indicating that the classical model is robust in this regime).

\section{Acknowledgments\label{section:ack}}
We thank S. L. Cornish, A. S. Bradley, T. P. Billam, D. I. H. Holdaway, P. Mason and A. L. Marchant for useful discussions and the UK EPSRC (grant numbers EP/G056781/1 and EP/K03250X/1 ) and The Royal Society (grant no. IE110202) for support. S.J.R. is supported by the University of Otago, and thanks Durham University for their hospitality.


%
\end{document}